# Research Activity Classification based on Time Series Bibliometrics

Takahiro Kawamura[*], Yasuhiro Yamashita[*] and Katsuji Matsumura[*]

[*] takahiro.kawamura@jst.go.jp; yasuhiro.yamashita@jst.go.jp; matsumur@jst.go.jp
Japan Science and Technology Agency, Tokyo (Japan)

## INTRODUCTION

In this paper, we aim to discover "distinguished" researchers who have not been widely known based on bibliometrics. In many cases, some criteria concerning the amount of research achievements, such as the number of papers published and citations received, are determined, and then the research activities are evaluated regarding whether the criteria are satisfied. However, the simple sum of the achievements is consequently beneficial to the elder researchers, and there is a problem in that researchers who have a significant achievement cannot be differentiated from those publishing a few papers over a long period of time. Also, it has been found that the transitions of the achievements of "distinguished" researchers exhibit some patterns (Bjork, Offer & Soederberg 2014). Therefore, this paper finds characteristic patterns from the time series changes of the international and domestic research achievements of "distinguished" researchers, and then it attempts to classify the researchers.

## CLASSIFICATION USING TIME SERIES METRICS
### Feature Generation

There are several ways to represent time series data as features, such as numerical values and item pairs of an attribute and its value. In this paper, we convert the sequential data of real values to characters in order to reduce the data size. Symbolic Aggregate Approximation (SAX) (Patel et al, 2002) is a well-known method for this purpose. We used the SAX but converted the differences between the two values to represents the changes in achievements instead of the values of the sequential data in the SAX. We also used k-gram (consecutive k characters) in Natural Language Processing (NLP), and then extended it to the combined k-grams that have anteroposterior relations in time, to represent loosely the multiple overlapping sequences.

Figure 1 shows the workflow of feature generation. First, the numerical values are normalized to [0–100] per person, and then differences between consecutive years are converted to six symbols [U, u, S, d, D, 0] to represent changes in time as features, where U = over a 30 point increase from the last year, u = 5–0 point increase, S = +/-5 point change including no change, d = between -5 and -30 point decrease, D = over a -30 point decrease, and 0 = no paper/cited. Then, we generate a k-gram (k = 1–4) of the symbols, where k = 4 indicates a five-year period, and, finally, the k-grams for the five metrics in the following section are combined with anteroposterior relations (+, =, -), comparing the start time t of two time-series, where + indicates that the start time of the following k-gram is after the start time of the first k-gram, = indicates that the start time is the same as the first k-gram, and - indicates that the start time is before the first k-gram.

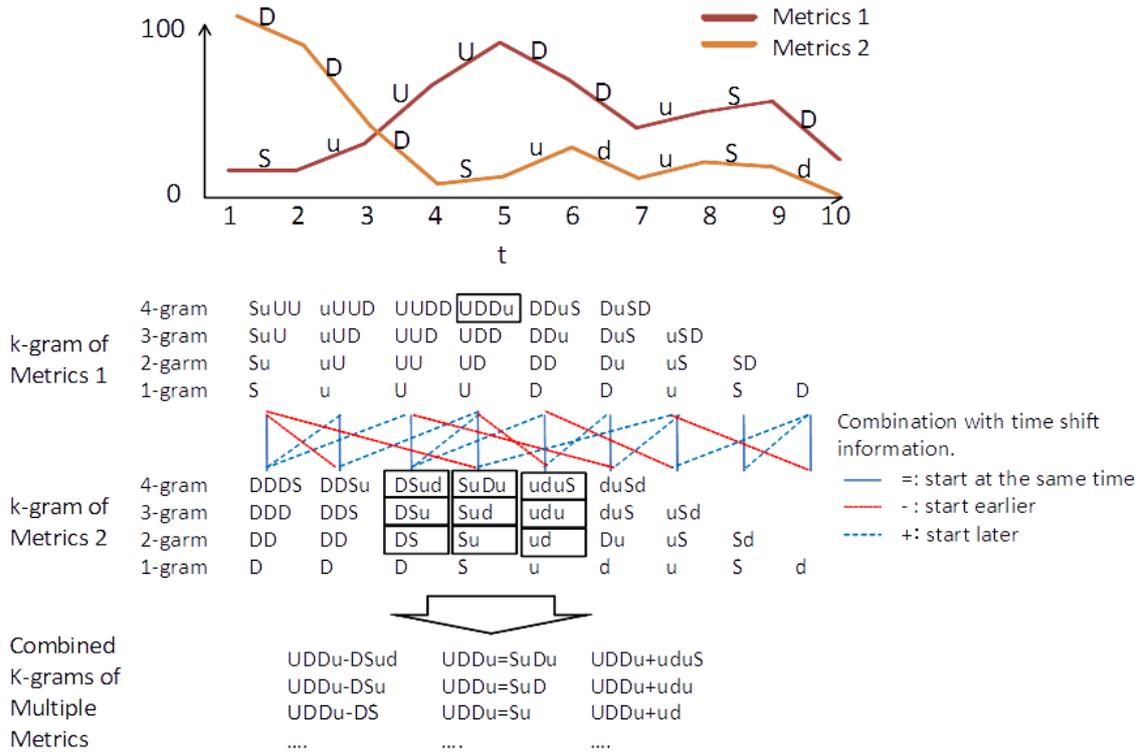

Figure 1: Feature Generation.

**Feature Selection**

Next, to find characteristic patterns from a set of patterns created in the previous section, which are important features in the following machine learning phase, we calculate $P_{ij}$ (Equation 1) for each pattern, which corresponds to Term Frequency/Inverse Document Frequency (TF/IDF) in NLP. $P_{ij}$ filters the general patterns that are common with several researchers. Finally, we selected the patterns of the ten highest $P_{ij}$ per person. Thus, if there are 50 persons, at most 500 patterns become features after deleting duplicated patterns.

$$P_{ij} = paf_{ij} \times log(\frac{N}{pef_j}) \qquad (1)$$
$$paf_{ij} = \# \text{ occurences of pattern } j \text{ in person } i$$
$$pef_j = \# \text{ persons containing pattern } j$$
$$N = \# \text{ persons in total}$$

**EXPERIMENT ON RESEARCHER CLASSIFICATION**
**Summary of Experiment Dataset**
In the experiment, we used the bibliographic datasets of the JST and Elsevier's Scopus (http://www.scopus.com). The JST dataset includes 5,000 international journal titles and 9,500 domestic journal titles for science and technology. The Scopus dataset includes 21,000 international journal titles, 417 domestic (Japanese) journal titles. We then retrieved sequential data concerning the following five metrics by year.

・   # of papers from domestic journals and conferences
・   # of papers from international journals and conferences
・   # of citations in domestic papers

- # of citations in international papers
- % of the first author's papers in the total of domestic and international papers

The datasets of researchers include 42 who specialize in Artificial Intelligence (AI) and 72 who specialize in Bioscience (Bio) in Japan (to the best of our knowledge, the order of authors in papers is not alphabetical in those domains). We randomly collected researchers belonging to The Japanese Society for Artificial Intelligence (http://www.ai-gakkai.or.jp/en) and The Japanese Society for Regenerative Medicine (https://www.jsrm.jp/?lang=english). Then, researchers who had received a grant of more than 30 million JPN as the project representative were labelled as "distinguished" (TRUE in the classification) by referring to the Database of Grants-in-Aid for Scientific Research (https://kaken.nii.ac.jp/en/), since grants are given to "distinguished" researchers who propose excellent themes and are determined to achieve those themes through the sufficient deliberation of several domain experts. Sequential data in time were taken for ten years before receipt of the grant, since the grant necessarily increases research achievements. For researchers who do not have such grants (FALSE in the classification), the sequential data were taken in the last decade leading to 2014. The number of researchers with a FALSE classification is much larger than the number of them with TRUE. However, since we assume to have a screening process conducted before the proposed method, the distribution of researchers is set to be equal to each other. Thus, we selected the same number of researchers with and without grants (TRUE and FALSE).

**Classification accuracy**

This section first presents a baseline result based on the sum of data for ten years concerning the above four metrics and the average for the first author ratio. The features are the numerical values of the achievements instead of their changes in time (k-gram). Table 1 (above) shows the accuracy of researcher classification (TRUE, FALSE) by the 10-fold cross validation using a decision tree, in which the algorithm is C4.5.

Table 1. Classification accuracy (%).

| Domain and Features | Class | Precision | Recall | F-measure |
|---|---|---|---|---|
| AI by quantity | TRUE | 80.0 | 76.2 | 78.0 |
|  | FALSE | 77.3 | 81.0 | 79.1 |
| Bio by quantity | TRUE | 64.2 | 89.5 | 74.7 |
|  | FALSE | 82.6 | 50.0 | 62.3 |
| AI by time series and quantity | TRUE | 95.0 | 90.5 | 92.7 |
|  | FALSE | 90.9 | 95.2 | 93.0 |
| Bio by time series and quantity | TRUE | 87.1 | 75.0 | 80.6 |
|  | FALSE | 78.0 | 88.9 | 83.1 |

Next, we combined the feature vectors of the time series patterns representing the changes and the above five features in quantity. The result is shown in Table 1 (below), and thus we can confirm that the combination of both features has higher accuracy than the amount of the achievements alone. We also conducted the Chi-squared test for independence to assess the statistical significance $p$ between the numbers of correctly and incorrectly classified researchers in Tables 1. The results of the AI and Bio domains in Table 1 (below) were superior to those in Table 1 (above) ($p = 0.014$). Thus, there is a statistically significant difference in the classification between the features in quantity and the combination of the time series patterns with them.

**RELATED WORK**
There are few case studies on time series analysis, and most papers visualize time series changes in terms of specific metrics. For instance, Prathap (2011) proposed exergy as a single number scalar indicator based on a thermodynamic analogy in order to assess the bibliometrics progress of researchers and then represented the progress as a phase diagram. However, Leydesdorff (2013) argues that the sciences evolve as complex and non-linear systems that contain recursive terms and interaction, for example, between universities and industries. Multivariate analysis in bibliometrics has focused mainly on static designs and should address more of its dynamic developments. Bjork, Offer & Soederberg (2014) also proved that there are patterns in the transition of research achievements, as described in the introduction. In terms of the publications of 57 Nobel Prize winners in economics from 1930 to 2005, the study indicated that time series changes in the number of citations received can be classified into four types and also fit an innovation diffusion curve derived from the Bass model. Thus, Kajikawa et al. conducted a Topological Data Analysis of the citation networks of papers. In this study, time series changes in the position of specific papers in the network are represented by three measures of centrality: clustering centrality, closeness centrality, and betweenness centrality; then, the correlation with the number of citations that will be received in the future are estimated (Shibata, Kajikawa & Matsushima 2007), (Iwami et al. 2014). Although the approach is different, the purpose is the same as that of our study.

**CONCLUSION AND FUTUREWORK**
To provide useful reference information other than the simple sum of the metrics in the examination of research grants, this paper proposed a classification method for researchers based on time series bibliometrics. Future works include an increase in the number of researchers as well as verification in other domains.